\documentclass[10pt,conference]{IEEEtran}

\usepackage{dsfont}
\usepackage{amsmath}
\usepackage{amssymb}
\usepackage{color}
\usepackage{bbm}
\usepackage[dvips]{graphicx}
\usepackage{algorithmic}
\usepackage[]{algorithm}
\usepackage{centernot}

\graphicspath{{figures/}}
\usepackage[numbers,sort&compress]{natbib}

\newcommand{\operator}[1]{\mathrm{#1}}

\newcommand{\ve}[1]{{\mathbf{#1}}}

\newcommand{\e}{\operator{E}}
\newcommand{\D}{\operator{D}}
\newcommand{\p}{\operator{Pr}}

\newcommand{\vv}{{\ve{v}}}

\newtheorem{definition}{Definition}

\newtheorem{theorem}{Theorem}

\newtheorem{lemma}{Lemma}
\newtheorem{remark}{Remark}

\long\def\symbolfootnote[#1]#2{\begingroup%
 \def\thefootnote{\fnsymbol{footnote}}\footnote[#1]{#2}\endgroup} 

\newcounter{customlistcounter}
\newenvironment{customlist}[2]{
\setcounter{customlistcounter}{\value{#1}}
\begin{list}{#2}{\usecounter{#1}
\settowidth{\labelwidth}{#2}
\setlength{\leftmargin}{\labelwidth}
\addtolength{\leftmargin}{\labelsep}}
\setcounter{#1}{\value{customlistcounter}}
}{
\end{list}
}

\newcounter{assume}
\renewcommand{\theassume}{(A.\arabic{assume})}

\title{Strong Secrecy and Stealth for Broadcast Channels with Confidential Messages}

\author{Igor Bjelakovic, Jafar Mohammadi, and S\l awomir Sta\'nczak \\
Fachgebiet Informationstheorie und theoretische Informationstechnik\\
 Technische Universit\"at Berlin, Einsteinufer 27, 10587 Berlin, Germany\\
Email: i.bjelakovic@tu-berlin.de, \{jafar.mohammadi, slawomir.stanczak\}@hhi.fraunhofer.de\\
}

\begin{document}
\maketitle

\symbolfootnote[0]{We present detailed proofs in the extended versions of this work.}

\begin{abstract}
This paper extends the weak secrecy results of Liu et al.\ for 
broadcast channels with two confidential messages to strong secrecy. Our results are based on an extension of the techniques developed by Hou and Kramer on bounding 
Kullback-Leibler divergence in context of \textit{resolvability} and \textit{effective secrecy}.
\end{abstract}


\section{Introduction} 

Based on the pioneering work of Shannon \cite{ShannonSecrecy49}, Wyner \cite{ WynerWireTap75} determined the secrecy capacity of a class of wiretap channels.
Wyner's work has been generalized by  Csisz\'ar and K\"orner \cite{CsiszarKoernerBroadcastSecure78} to the non-degraded broadcast channel with a single confidential message for one user
and a common message intended for both users. The confidential message has to be kept secret from the other user, while both of them decode the common message.
A variant of Csisz\'ar and K\"orner's model with two confidential messages and no common message was first studied by Liu et al.\ in \cite{LiuYatesDMCBCCM2008} for a discrete memoryless channel 
and later for the Gaussian case in \cite{LiuVPoorBCCM2009}. An inner and outer bound on the secrecy region can be found in \cite{LiuYatesDMCBCCM2008}.

The secrecy criterion used in \cite{WynerWireTap75, CsiszarKoernerBroadcastSecure78,LiuYatesDMCBCCM2008}, is the \emph{normalized} mutual information between the message 
 and the output distribution of the user regarded as the eavesdropper.
 This security criterion, called weak secrecy, delivers a very restricted security against eavesdropping attacks as was shown by Maurer in \cite{Maurer94thestrong}.
An \emph{unnormalized} definition of secrecy, called strong secrecy, was proposed in \cite{Maurer94thestrong} and large parts of previous work was extended from weak to strong secrecy
 (e.g. \cite{Maurer94thestrong}, \cite{CsiszarRuss}, \cite{BlochSSR13}, and, etc.).
 In this paper we extend the inner bound of \cite{LiuYatesDMCBCCM2008} from weak to strong secrecy. Along the proof we extend the method of
 Hou and Kramer \cite{JiHouKramer2014} based on even stronger notion of \emph{effective secrecy} to this scenario. More precisely we prove, in addition to strong secrecy, a stealthy communication (i.e. the presence of meaningful communication is hidden) is possible.

\subsection{Notation} We use capital letters for random variables (RV). If $X$ is a RV then $x$ is used to refer to an observation of $X$. Sets are denoted by $\mathcal{X}$ and $\mathcal{P}(\mathcal{X})$
stands for the set of probability distributions defined on the (finite set) $\mathcal{X}$. For the RVs $X_1,\dots, X_K$ with values in $\mathcal{X}_1, \dots, \mathcal{X}_K$ we write $X_1-X_2-\dots-X_K $ if they form a Markov chain. 
The symbol $\mathbf{x}$ stands for $x^{n}:=(x_1,\ldots, x_n)\in \mathcal{X}^n$. 
The mutual information between the RV's $A$ and $B$ is denoted by $I(A,B)$, while $H(A)$ and $H(A|B)$ are entropy of $A$ and conditional entropy of $A$ given $B$, respectively. 
Probability mass function of $X$ is $P_X(x)$ or in short $P(x)$ while probability of an event $E$ is denoted by $\p(E)$. Kullback-Leibler divergence of two probability distributions $P, Q$ defined on set $\mathcal{A}$ is given by, 
\begin{align} \label{eq: KLdef}
\D(P\|Q):= \Bigg\{ \begin{matrix}
\sum_{a\in\mathcal{A}}  P(a)\log \frac{P(a)}{Q(a)} &  & \mathrm{if }\; P \ll Q \\
+\infty  & & \mathrm{if }\; P \centernot \ll Q
\end{matrix}
\end{align} 
where $P \ll Q$ means that $P(a)=0$ whenever $Q(a) = 0$ $a \in \mathcal{A}$. 
The typical set of sequences $x^n \in \mathcal{X}^n$ for RV $X$ and $\epsilon >0 $ is denoted by $\mathcal{T}^n_{\epsilon}(P_X)$ as defined in \cite{GamalNIT}. 
We will freely use the properties of typical sets from \cite{GamalNIT}.
%

\section{System Model} 
We consider a broadcast scenario consisting of a sender (S) and two receivers. 
We assume that all channels are discrete memoryless with finite input alphabet $\mathcal{X}$, and finite output 
alphabets $\mathcal{Y}_1$ and $\mathcal{Y}_2 $. 
The conditional probability distribution governing the discrete memoryless 
broadcast channel (DMBCC) is given by
%
\begin{align}
P(\mathbf{y}_1,\mathbf{y}_2|\mathbf{x}) = \prod_{i=1}^{n} P(y_{1i},y_{2i}|x_i), 
\end{align}
where, $\mathbf{x}=x^n \in \mathcal{X}^n $, $\mathbf{y}_t=y^n_t= (y_{t1}, \dots, y_{tn}) \in \mathcal{Y}_t^n $, and $t \in \{1,2\}$.\\
The stochastic encoder at the sender is defined to be, 
\begin{align}\label{eq: stoch-encoder}
f: \mathcal{M}_1
\times\mathcal{M}_2 \rightarrow \mathcal{P}(\mathcal{X}^n) 
\end{align}
with,
\begin{align}
\sum_{\mathbf{x}\in \mathcal{X}^n} f(\mathbf{x}|m_1,m_2) = 1 \;\; \forall m_1 \in \mathcal{M}_1,\; m_2 \in \mathcal{M}_2  
\end{align}
where $\mathcal{M}_1 := \{1, \dots, M_1\}$ and $\mathcal{M}_2 := \{1, \dots, M_2\}$ are the message sets for receiver 1 and 2, respectively. 
%
%
%
%
%
The decoder at the $t$th node, $t \in \{1,2\}$, is defined as
\begin{align} 
g_t: \mathcal{Y}_t^n \rightarrow \mathcal{M}_t.
\end{align}
\begin{definition}[Strong Secrecy \cite{Maurer94thestrong}]
For every $\epsilon>0$ there is a non-negative integer $N(\epsilon)$ such that for all $n \geq N(\epsilon)$, 
\begin{align} \label{eq: equevo1}
I(W_1;Y^n_2|W_2) \leq \epsilon \\ 
I(W_2;Y^n_1|W_1) \leq \epsilon \label{eq: equevo2}
\end{align} 
where the RVs $W_1$ and $W_2$ are distributed uniformly over $\mathcal{M}_1$
and $\mathcal{M}_2$, respectively and the mutual information values are computed with respect to the distribution
\begin{align*} 
 P_{W_{1}W_{2}Y_{1}^{n}Y_{2}^{n}}(m_1, &m_2, \mathbf{y}_1, \mathbf{y}_2)= \\
&\frac{1}{M_1}\frac{1}{M_2}\sum_{\mathbf{x} \in \mathcal{X}^n} f(x^n|m_1, m_2)P(\mathbf{y}_1, \mathbf{y}_2|\mathbf{x}),
\end{align*}
with the stochastic encoder given in (\ref{eq: stoch-encoder}). 
\end{definition}

The probability of error at each node $t \in \{1,2\}$ is 
\begin{align*}
P^n_{e\;t} = \frac{1}{M_1M_2} \sum_{(m_1, m_2) \in \mathcal{M}_1 \times \mathcal{M}_2 } P[g_t(y^n_t) \neq m_t| (m_1,m_2) \;\mathrm{is\; sent}],
\end{align*}
\begin{definition}\label{def: achieve}
A rate pair ($R_1,R_2$) is said to be achievable, if for every $\epsilon' >0$, $\epsilon >0$, $\delta >0$, there exists a code ($M_1, M_2, n, P^n_{e \;1}, P^n_{e \;2}$)  for sufficiently large $n$
,in addition to (\ref{eq: equevo1}) and (\ref{eq: equevo2}), we have
\begin{align*}
P^n_{e\;t} \leq \epsilon' \;\; \textrm{ and }\;\; \frac{1}{n}\log M_t\ge R_t-\delta, \;\;\; t\in \{1,2\}. 
 \end{align*}
\end{definition}

\begin{remark}The above definition provides a \textit{strong} condition on security. More details are given in section \ref{sec: strongseccriterion} where we show that even stronger notion of secrecy based on stealth can be achieved. 
\end{remark}
\section{Strong Secrecy for Broadcast Channels}
In this section we present an achievable secure rate region with strong secrecy criterion for a DMBCC. The following theorem summarizes our results: 
\begin{theorem}
The rate pair ($R_1,R_2$) is achievable, in the sense of Definition \ref{def: achieve}, for DMBCC with confidential messages and strong secrecy criterion, if   
\begin{align*}
0 \leq& R_1,\;0 \leq R_2\\
R_1 <& I(V_1;Y_1|U)-I(V_1;Y_2|V_2,U) - I(V_1;V_2|U) \\
R_2 <& I(V_2;Y_2|U)-I(V_2;Y_1|V_1,U) - I(V_1;V_2|U)
\end{align*}
where the information quantities are computed with respect to some probability distributions $P_{U,V_{1},V_{2},X ,Y_1 ,Y_2}$ such that Markov chain condition $U-(V_1, V_2)-X-(Y_1, Y_2)$ holds and $X$ and $(Y_1, Y_2)$
are connected via the given broadcast channel. The auxiliary RVs $U,V_1$ and $ V_2$ take values in finite sets $\mathcal{U}, \mathcal{V}_1$ and $ \mathcal{V}_2$. \footnote{The cardinalities of $\mathcal{U}, \mathcal{V}_1, \mathcal{V}_2$ can be bounded by the Ahlswede-K\"otner technique. We shall provide this in the journal version of this manuscript} 
\end{theorem}  

\begin{IEEEproof}
The proof, which consists of two parts, achievability and secrecy, unfolds in the following subsections. 
The achievablity proof is based on techniques developed in \cite{marton79, gelfandpinsker} and extends those devoted to secrecy 
developed in \cite{JiHouKramer2014} for the wiretap channel to the present setting.  

\subsection{Coding scheme}
\begin{remark}
 In order to simplify our notation we will drop the auxiliary random variable $U$ in the following proof. It can be included into the proof via standard arguments \cite{CsiszarKornerBook, ElGamalMeulen81}.
\end{remark}
For each $m_t \in \{1, \dots, 2^{nR_t}\},
s_t \in \{1,\dots, 2^{nR'_t}\}$, and $
k_t \in \{1, \dots, 2^{nR_{co}}\}$, $t\in \{1,2\}$, we draw independently sequences $\mathbf{v}_t(m_t,s_t,k_t) $ according to 
$P^n(\mathbf{v}_t) = \prod_{i=1}^{n}P(v_{t,i})$.\\
Let $0<\epsilon<\epsilon'<\epsilon''$. For $(m_1,m_2)$ and $(s_1,s_2)$ find a pair $(k_1,k_2)$ such that
\begin{align*}
 (\mathbf{v}_1(m_1,s_1,k_1),\mathbf{v}_2(m_2,s_2,k_2)) \in \mathcal{T}^n_{\epsilon}(P_{V_1 V_2}).
\end{align*}
If there is no such a pair choose $(k_1,k_2)=(1,1)$. Then select a sequence $\mathbf{x}(m_1,m_2,s_1,s_2)\in \mathcal{X}^n$ with
\begin{align*}
 &(\mathbf{v}_1(m_1,s_1,k_1),\mathbf{v}_2(m_2,s_2,k_2), \mathbf{x}(m_1,m_2,s_1,s_2)) \\ 
 & \hspace{6cm} \in \mathcal{T}^n_{\epsilon'}(P_{V_1 V_2 X}).
\end{align*}

\subsubsection*{Encoding} To transmit $(m_1,m_2)$ select uniformly at random a pair $(s_1,s_2)$ and send $ \mathbf{x}(m_1,m_2,s_1,s_2)$.

\subsubsection*{Decoding} Upon receiving $\mathbf{y}_t$ decoder, $t\in \{ 1,2 \}$, declares $(m_t,s_t)$ is sent if it is unique pair such that
\begin{align}
(\mathbf{v}_t(m_t,s_t,k_t), \mathbf{y}_t) \in \mathcal{T}^n_{\epsilon''}(P_{V_t Y_t})
\end{align}

\subsection{Error Analysis} 
The error analysis is carried out using standard arguments. Using the mutual covering lemma, packing lemma, and the properties of the typical sequences (cf. for example \cite{GamalNIT})
we obtain
\begin{align} \label{eq: R_coCoverLem}
R_{co} > I(V_1;V_2)\quad\textrm{ and }\quad R_1 + R'_1 + R_{co} &< I(V_1;Y_1).
\end{align}
and
\begin{align*}
\lim_{n\to \infty}P^n_{e\;t} =0\qquad t\in \{1,2\} 
 \end{align*}
 exponentially fast.\\
From (\ref{eq: R_coCoverLem}) we have 
\begin{align} \label{eq: ratReleible1}
R_1 + R'_1 &< I(V_1;Y_1) - I(V_1;V_2)
\end{align}
and by symmetry, 
\begin{align} \label{eq: ratReleible2}
R_2 + R'_2 &< I(V_2;Y_2) - I(V_1;V_2).
\end{align}
The bounds on $R_1 '$ and $R_2'$ are derived in the following subsection.
\subsection{Strong Secrecy Criterion} \label{sec: strongseccriterion}
\begin{remark}
For the secrecy analysis we drop the random variable $U$ for the sake of notational simplicity. 
It can be introduced, if desired via standard arguments at the end of the proof \cite{CsiszarKoernerBroadcastSecure78, JiHouKramer2014}.
\end{remark}

Before we further continue, we define a deterministic function $\phi_{\tau}$ on pairs 
$\{ \vv_1(m_1,s_1,k_1) \}_{k_1=1}^{2^{nR_{co}}} \times  \{\vv_2(m_2,s_2,k_2)\}_{k_2=1}^{2^{nR_{co}}}$ that returns a pair ($k_1,k_2$) such that 
\begin{align}
(\vv_1(m_1,s_1,k_1),\vv_2(m_2,s_2,k_2)) \in \mathcal{T}^n_{\epsilon}(P_{V_1,V_2}).
\end{align}
In the case that there are many such pairs, we choose one arbitrarily. However, if there are no 
such pairs, the function $\phi_\tau$ returns (1,1). From now on, we denote the codeword pairs simply 
$(\vv_1(m_1,s_1), \vv_2(m_2,s_2))$ based on selection function $\phi_\tau $.

We extend the framework in \cite{JiHouKramer2013, JiHouKramer2014} to find a condition that bounds 
\begin{align} \label{eq: KLbound}
\D(P_{W_1,Y_2|\mathbf{v}_2} ||P_{W_1}Q^n_{Y_2|\mathbf{v}_2}) \leq \xi
\end{align}
where the left hand side is known as the\textit{ effective secrecy},    
$\xi >0$ is arbitrarily small and 
\begin{align*} 
P_{W_1, Y_2| \mathbf{v}_2}(m_1, & \mathbf{y}_2| \vv_2) : = \\
&\frac{1}{M_1} \sum_{s_1=1}^{J_1} \frac{1}{J_1} P^n(\mathbf{y}_2|\mathbf{v}_1(m_1,s_1),\vv_2(m_2,s_2)),
\end{align*}
and further $Q^n_{Y_2|\vv_2}$ is distribution of $Y^n_2$ given $\vv_2$ while no meaningful 
message is transmitted for $W_1$,
\begin{align} \label{eq: stealthdistro}
Q^n_{Y_2|V_2}(\mathbf{y}_2| \vv_2) = \sum_{\vv_1} P^n_{V_1}(\vv_1) P^n_{Y_2|V_1,V_2}(\mathbf{y}_2|\vv_1, \vv_2)
\end{align}
 and finally, $P_{W_1}$ is distribution of $W_1$. We can further expand (\ref{eq: KLbound}) as  
\begin{align} \nonumber
\D(&P_{W_1,Y_2|\vv_2} ||P_{W_1}Q^n_{Y_2|\vv_2}) \\ \nonumber
= &\sum_{m_1,\mathbf{y}_2} P_{W_1,Y_2|\vv_2}(m_1, \mathbf{y}_2| \vv_2) \\ \nonumber
& \hspace{3.8cm} \log \bigg( \frac{P_{W_1,Y_2|\mathbf{v}_2}(m_1, \mathbf{y}_2| \vv_2) }{Q^n_{Y_2|\mathbf{v}_2}(\mathbf{y}_2| \vv_2)P_{W_1}(m_1)} \bigg) \\ \nonumber
=&\sum_{m_1,\mathbf{y}_2} P_{W_1,Y_2|\vv_2}(m_1, \mathbf{y}_2| \vv_2) \\ \nonumber
& \hspace{1.4cm}  \log \bigg( \frac{P_{W_1,Y_2|\mathbf{v}_2}(m_1, \mathbf{y}_2| \vv_2) }{Q^n_{Y_2|\mathbf{v}_2}(\mathbf{y}_2| \vv_2)P_{W_1}(m_1)} \cdot
\frac{P_{Y_2|\vv_2}(\mathbf{y}_2| \vv_2)}{P_{Y_2|\vv_2}(\mathbf{y}_2| \vv_2)} \bigg) 
\\ \nonumber
=&\sum_{m_1,\mathbf{y}_2} P_{W_1,Y_2|\vv_2}(m_1,\mathbf{y}_2| \vv_2) \\ \nonumber
&  \hspace{1.4cm} \log \bigg( \frac{P_{W_1,Y_2|\mathbf{v}_2}(m_1, \mathbf{y}_2| \vv_2) }{P_{W_1}(m_1)P_{Y_2|\vv_2}(\mathbf{y}_2| \vv_2)} 
\cdot \frac{P_{Y_2|\vv_2}(\mathbf{y}_2| \vv_2)}{Q^n_{Y_2|\mathbf{v}_2}(\mathbf{y}_2| \vv_2)} \bigg) 
\\ \nonumber
\overset{(\alpha)}{=} & \sum_{m_1,\mathbf{y}_2} P_{W_1,Y_2|\vv_2}(m_1, \mathbf{y}_2| \vv_2)  \log \bigg( \frac{P_{W_1,Y_2|\mathbf{v}_2}(m_1, \mathbf{y}_2| \vv_2) }{P_{W_1}(m_1)P_{Y_2|\vv_2}(\mathbf{y}_2| \vv_2)}\bigg)  \\ \nonumber
& \hspace{2.1cm} + \sum_{\mathbf{y}_2} P_{Y_2|\vv_2}( \mathbf{y}_2| \vv_2)  \log \bigg(\frac{P_{Y_2|\vv_2}(\mathbf{y}_2| \vv_2)}{Q^n_{Y_2|\mathbf{v}_2}(\mathbf{y}_2| \vv_2)} \bigg) \\
=& I(W_1; Y^n_2|\mathbf{v}_2) + 
\D(P_{Y_2|\mathbf{v}_2} ||Q^n_{Y_2|\mathbf{v}_2}),\label{eq: stealthandsecrecy}
\end{align}
in $(\alpha)$ we used that $P_{Y_2|V_2}(\mathbf{y}_2|\vv_2)=\sum_{m_1}P(m_1, \mathbf{y}_2| \vv_2)$. 

\begin{remark} The expansion in (\ref{eq: stealthandsecrecy}), reveals that (\ref{eq: KLbound}) 
addresses not only the strong secrecy notation by bounding $I(W_1; Y^n_2|\mathbf{v}_2)$, 
but also bounds 
$\D(P_{Y_2|\mathbf{v}_2} ||Q^n_{Y_2|\mathbf{v}_2})$ which corresponds to \textit{stealth} of the communication. 
\end{remark}
%
%
Based on chain rule for informational divergence we have, 
\begin{align} \label{eq: chainruleDivDis}
&\D(P_{W_1,Y_2|\vv_2} ||P_{W_1}Q^n_{Y_2|\mathbf{v}_2})  = \\
&\hspace{1cm} \D(P_{W_1} ||P_{W_1}) + \D(P_{Y_2|\vv_2,W_1} ||Q^n_{Y_2|\mathbf{v}_2}|P_{W_1}), \nonumber
\end{align}
where, according to the definition of $\D(\cdot\|\cdot)$ in (\ref{eq: KLdef}), the first term above equals to zero, thus we proceed with bounding  $\D(P_{Y_2|\vv_2,W_1} ||Q^n_{Y_2|\mathbf{v}_2}|P_{W_1})$. 
The following lemma provides an upperbound on $\D(P\|Q)$, which is useful for the rest of the proof. 
\begin{lemma}\label{lem: KLupperbound}
For probability distributions $P$ and $Q$, defined on a finite set $A$, with $P \ll Q$, we have, 
\begin{align} \label{eq: InfoDivergUB}
\D(P\|Q) \leq \log \frac{1}{\pi_Q}
\end{align} 
where, $\pi_Q = \min\{ Q(a):Q(a)>0\}$.
\end{lemma}
Taking expectation with respect to $(V^n_1,V^n_2)$ of  $\mathbbm{1}{(\phi_\tau = (1,1))} \D(P_{Y_2|V_2,W_1} ||Q^n_{Y_2|V_2}|P_{W_1})$  yields,
\begin{align} \label{eq: functiontypicality}
&\e \big[ \mathbbm{1}{(\phi_\tau = (1,1))} \D(P_{Y_2|V_2,W_1} ||Q^n_{Y_2|V_2}|P_{W_1})   \big] \\ \nonumber & \hspace{2.8cm} \leq \p\big(\phi_\tau=(1,1)\big)\log\frac{1}{(\pi_{Q_{Y_2|V_2}})^n} \\\nonumber
& \hspace{2.8cm}=  \p\big(\phi_\tau=(1,1)\big)(-n)\log(\pi_{Q_{Y_2|V_2}})\\ 
& \hspace{2.8cm} \leq 2^{-n\alpha} \label{eq: functiontypicality1}
\end{align}
for all sufficiently large $n$, where $\mathbbm{1}(\cdot)$ is an indicator function and we apply lemma \ref{lem: KLupperbound} in (\ref{eq: functiontypicality}). The last inequality comes from the mutual covering lemma (cf. \cite{GamalNIT} for example),
where $\alpha >0$ 
is a constant independent of $n$. With (\ref{eq: chainruleDivDis}) and (\ref{eq: functiontypicality1}), therefore, we have for all sufficiently large $n$ 
\begin{align} \label{eq: phi_typicalpart}
&\e[\D(P_{Y_2|W_1,\vv_2} ||Q_{Y_2|\mathbf{v}_2}|P_{W_1})]  \leq \\ \nonumber
 &\hspace{1.0cm} \e \big[ \mathbbm{1}{(\phi_\tau \neq (1,1))} \D(P_{Y_2|\vv_2,W_1} ||Q_{Y_2|\mathbf{v}_2}|P_{W_1}) + 2^{-n\alpha}. 
\end{align}
Hence, in the following we focus on bounding the first term in the right hand side of (\ref{eq: phi_typicalpart}).

We take the expectation over the $Y^n_{2}$, $V_1$, $W_1$ and $S_1$; we obtain
\begin{align}  \nonumber
&\e[\D(P_{Y_2|W_1,\vv_2} ||Q^n_{Y_2|\mathbf{v}_2}|P_{W_1})] \\ \nonumber
 &= \e \bigg[ \log \frac{\sum_{s_1=1}^{J_1} P(Y^n_2|V^n_1(W_1,s_1),\vv_2)}{J_1 Q^n_{Y_2|\mathbf{v}_2}(Y^n_2|\vv_2)} \bigg] \\  \nonumber
&= \sum_{m_1,s_1} \frac{1}{J_1 M_1} \e \bigg[ \log \frac{\sum_{i=1}^{J_1} P(Y^n_2|V^n_1(m_1,i),\vv_2)}{J_1 Q^n_{Y_2|\mathbf{v}_2}(Y^n_2|\vv_2)} \bigg| \\ \nonumber
& \hspace{6cm} W_1=m_1, S_1=s_1 \bigg]\\ \label{eq: upperKLjensen}
&\leq \sum_{m_1,s_1} \frac{1}{J_1 M_1} \e \bigg[ \log \frac{P(Y^n_2|V^n_1(m_1,s_1),\vv_2)}{J_1 Q^n_{Y_2|\mathbf{v}_2}(Y^n_2|\vv_2)}  \\ \nonumber
& \hspace{4.4cm} +\frac{J_1-1}{J_1}\bigg| W_1=m_1, S_1=s_1 \bigg]\\ \nonumber
&\leq \sum_{m_1,s_1} \frac{1}{J_1 M_1} \e \bigg[ \log \frac{P(Y^n_2|V^n_1(m_1,s_1),\vv_2)}{J_1 Q^n_{Y_2|\vv_2}(Y^n_2|\vv_2)} +1\bigg| \\ \nonumber
& \hspace{6cm} W_1=m_1, S_1=s_1 \bigg]\\ 
&= \e \bigg[ \log \bigg( \frac{P(Y^n_2|V^n_1,\vv_2)}{J_1 Q^n_{Y_2|V_2}(Y^n_2|\vv_2)} +1 \bigg) \bigg] \label{eq: upperKL},
\end{align}
where, in (\ref{eq: upperKLjensen}) we used Jensen's inequality applied to the part of the expectation over $V^n_1(m_1,i)$ for $i \neq s_1$.
\footnote{A detailed proof of this derivation is due to journal version of this work}. 

Before proceeding with the bounds on the informational divergence, we introduce the following lemma. 
\begin{lemma} \label{lem: typicalPenalty}
Let $P_{V_1,V_2, Y_2}(y_2,v_2,v_1) = P_{V_1,V_2}(v_1,v_2)P_{Y_2|V_1,V_2}(y_2|v_1,v_2) $
be a probability distribution on $\mathcal{Y}_2\times\mathcal{V}_1\times \mathcal{V}_2$.
For $\epsilon \in (0,\frac{1}{2})$ and distribution 
\begin{align}
Q(y_2|v_2) := \sum_{v_1} P_{V_1}(v_1)P_{Y_2|V_1,V_2}(y_2|v_1,v_2)
\end{align} 
it holds for $(\vv_2,\mathbf{y}_2) = \mathcal{T}^n_\epsilon (P_{V_2,Y_2})$ that
\begin{align}
Q^n(\mathbf{y}_2|\vv_2) \geq 2^{-n(H(Y_2|V_2) + I(V_1;V_2)+ \delta(\epsilon))}
\end{align}
with $\delta(\epsilon)>0$ and $\lim_{\epsilon \rightarrow 0}\delta(\epsilon) = 0 $.  
\end{lemma}

\begin{IEEEproof}
According to Lemma 2.6 in \cite{CsiszarKornerBook} we have for all $(Y_2,V_2)$, 
\begin{align} \label{eq: lemtypPENALTY}
Q^n(\mathbf{y}_2|\vv_2) = 2^{-n(H(P^e_{V_2,Y_2}) - H(P^e_{V_2}) + \D( P^e_{V_2,Y_2}\|P^e_{V_2}Q))}
\end{align}
where, 
\begin{enumerate}
\item $P^e_{V_2,Y_2}$ denotes the empirical distribution or type of the sequence pair $(\vv_2,\mathbf{y}_2)$
\item $P^e_{V_2}$ is the empirical distribution generated by the sequence $\vv_2$ 
\item the distribution $P^e_{V_2}Q$ is the joint distribution computed with respect to $P^e_{V_2}$ and $Q$
\end{enumerate}
Uniform continuity of the entropy (Lemma 2.7 in \cite{CsiszarKornerBook}), $(\vv_2,\mathbf{y}_2) = \mathcal{T}^n_\epsilon (P_{V_2,Y_2}) $, and the fact that $\epsilon \in (0,\frac{1}{2})$, yield
\begin{align} \label{eq: lemtypABS}
| H(P^e_{V_2,Y_2}) - H(P^e_{V_2}) - H(Y_2|V_2)| \leq \delta_1(\epsilon),
\end{align}
where, $\delta_1(\epsilon)>0$ with $\lim_{\epsilon \rightarrow 0}\delta_1(\epsilon)=0$.
Moreover, 
\begin{align} \label{eq: lemtypABS2}
| \D( P^e_{V_2,Y_2}\|P^e_{V_2}Q) - \D( P_{V_2,Y_2}\|P_{V_2}Q)| \leq \delta_2(\epsilon),
\end{align}
with $\delta_2(\epsilon) > 0 $ and $\lim_{\epsilon \rightarrow 0}\delta_2(\epsilon)=0$.\\
Inequality (\ref{eq: lemtypABS2}) can be seen as follows: In a first step we show that
$ P^e_{V_2,Y_2}\ll P^e_{V_2}Q $. To this end we assume
\begin{align*}
P^e_{V_2}Q(v_2, y_2)=P^e_{V_2}(v_2)Q(y_2|v_2) =0.
\end{align*}
Then either $P^e_{V_2}(v_2)=0$ and then $P^e_{V_2,Y_2}(v_2,y_2)=0 $ automatically.
Or $Q(y_2|v_2)=0$ and then
\begin{align}
 0=& Q(y_2|v_2)\\ \nonumber
 =& \sum_{v_1} P_{V_1}(v_1)P_{Y_2|V_1,V_2}(y_2|v_1,v_2)\\ \nonumber
 \ge & \sum_{v_1} P_{V_1, V_2}(v_1,v_2) P_{Y_2|V_1,V_2}(y_2|v_1,v_2)\\ \nonumber
 =& \sum_{v_1} P_{V_1,V_2, Y_2}(v_1,v_2,y_2)\\ \nonumber
 =& P_{V_2, Y_2}(v_2,y_2),
 \end{align}
which, again, implies $P^e_{V_2,Y_2}(v_2,y_2)=0 $ by definition of typical sequences, and thus $ P^e_{V_2,Y_2}\ll P^e_{V_2}Q $.
Since $ P^e_{V_2,Y_2}\ll P^e_{V_2}Q $ we have
\begin{align}
\D( P^e_{V_2,Y_2}\|P^e_{V_2}Q)\\ \nonumber
= & \sum_{v_2, y_2} P^e_{V_2, Y_2}(v_2,y_2)\log P^e_{V_2, Y_2}(v_2,y_2)\\ \nonumber
 & -\sum_{v_2, y_2} P^e_{V_2, Y_2}(v_2,y_2)\log P^e_{V_2}(v_2)\\ \nonumber
 & -\sum_{v_2, y_2} P^e_{V_2, Y_2}(v_2,y_2)\log Q(y_2|v_2)
\end{align}
and applying again the uniform continuity of entropy along with the properties of
typical sequences we obtain (\ref{eq: lemtypABS2}).\\
To proceed with the proof we need to show, 
\begin{align} \label{eq: lemtypmutualinfo}
 \D( P_{V_2,Y_2}\|P_{V_2}Q) \leq I(V_1,V_2).
\end{align}
This inequality can be derived as follows, 
\begin{align} \nonumber
 \D( &P_{V_2,Y_2} \|P_{V_2}Q)  \\ \nonumber
 =&\sum_{a,b} P_{V_2,Y_2}(a,b)\log\frac{P_{V_2,Y_2}(a,b)}{P_{V_2}(a)\sum_{c}P_{V_1}(c)P_{Y_2|V_1,V_2}(b|c,a)} \\ \nonumber
=& \sum_{c,a,b} P_{V_1,V_2,Y_2}(c,a,b)\log\frac{\sum_c P_{V_1,V_2,Y_2}(c,a,b)}{\sum_{c}P_{V_2}(a)P_{V_1}(c)P_{Y_2|V_1,V_2}(b|c,a)} \\ \label{eq: lemtyp1}
\leq & \sum_{c,a,b} P_{V_1,V_2,Y_2}(c,a,b)\log\frac{ P_{V_1,V_2,Y_2}(c,a,b)}{P_{V_2}(a)P_{V_1}(c)P_{Y_2|V_1,V_2}(b|c,a)} \\  \nonumber
= & \sum_{c,a,b} P_{V_1,V_2,Y_2}(c,a,b)\log\frac{ P_{V_1,V_2}(c,a)P_{Y_2|V_1,V_2}(b|c,a)}{P_{V_2}(a)P_{V_1}(c)P_{Y_2|V_1,V_2}(b|c,a)} \\ \nonumber
= & \sum_{c,a,b} P_{V_1,V_2,Y_2}(c,a,b)\log\frac{ P_{V_1,V_2}(c,a)}{P_{V_2}(a)P_{V_1}(c)} \\ \nonumber
= & \sum_{c,a} P_{V_1,V_2}(c,a)\log\frac{ P_{V_1,V_2}(c,a)}{P_{V_2}(a)P_{V_1}(c)} \\
=&  I(V_1,V_2), \label{eq: lemtyp2}
\end{align}
where, (\ref{eq: lemtyp1}) is by the Log-Sum-Inequality (Lemma 3.1 in \cite{CsiszarKornerBook}).

Combining (\ref{eq: lemtypABS}), (\ref{eq: lemtypABS2}), and (\ref{eq: lemtyp2}) 
leads to the claim of the lemma. 
\end{IEEEproof}

The expression (\ref{eq: upperKL}) can be split up as follows: (Recall that the expectation is taken over those codewords with $\phi_\tau\neq(1,1)$   
\begin{align*}
& \sum_{\substack{ \vv_1 \\ (\vv_1,\vv_2) \in \mathcal{T}^n_{\epsilon} }} P^n_{V_1}(\vv_1) 
 \sum_{\mathbf{y}_2} P^n_{Y_2|V_1,V_2}(\mathbf{y}_2|\vv_1, \vv_2) \\
 & \hspace{2cm} \log \bigg( \frac{P^n_{Y_2|\vv_2,\vv_1}(\mathbf{y}_2|\vv_2,\vv_1)}{J_1Q^n_{Y_2|\vv_2}(\mathbf{y}_2|\vv_2)} & +1 \bigg) = e_1 + e_2
\end{align*}
where,
\begin{align} \nonumber
e_1 := & \sum_{\substack{ \vv_1 \\ (\vv_1,\vv_2) \in \mathcal{T}^n_{\epsilon} }} P^n_{V_1}(\vv_1)
\sum_{ \substack{ \mathbf{y}_2 \\ (\vv_1, \vv_2, \mathbf{y}_2) \notin \mathcal{T}^n_\epsilon}} P^n_{Y_2|V_1,V_2}(\mathbf{y}_2|\vv_1, \vv_2) \\ 
&  \hspace{3cm} \log \bigg( \frac{P^n(\mathbf{y}_2|\vv_2,\vv_1)}{J_1Q^n(\mathbf{y}_2|\vv_2)} +1 \bigg) \label{eq: nontypicalset} \\ \nonumber
e_2 :=& \sum_{\substack{ \vv_1 \\ (\vv_1,\vv_2) \in \mathcal{T}^n_{\epsilon} }} P^n_{V_1}(\vv_1) 
\sum_{ \substack{ \mathbf{y}_2 \\ (\vv_1, \vv_2, \mathbf{y}_2) \in \mathcal{T}^n_\epsilon}} P^n_{Y_2|V_1,V_2}(\mathbf{y}_2|\vv_1, \vv_2) \\ 
& \hspace{3cm} \log \bigg( \frac{P^n(\mathbf{y}_2|\vv_2,\vv_1)}{J_1Q^n(\mathbf{y}_2|\vv_2)} +1 \bigg) \label{eq: typicalset} 
\end{align}
We upperbound $e_1$ in (\ref{eq: nontypicalset}) as
\begin{align} \nonumber
e_1 &\leq \sum_{\substack{ \vv_1 \\ (\vv_1,\vv_2) \in \mathcal{T}^n_{\epsilon} }} P^n_{V_1}
(\vv_1)\sum_{ \substack{ \mathbf{y}_2 \\ (\vv_1, \vv_2, \mathbf{y}_2) \notin 
\mathcal{T}^n_\epsilon}} \\ \nonumber
& \hspace{2cm}  P^n_{Y_2|V_1,V_2}(\mathbf{y}_2|\vv_1, \vv_2) \log \bigg( (\frac{1}{\pi_{Y_2|V_2}})^n + 1 \bigg) \\
&\leq  2|\mathcal{Y}_2| |\mathcal{V}_1| |\mathcal{V}_2|2^{-2n\epsilon^2\pi^2_{(Y_2|V_1,V_2)}}(-n)\log  (\pi_{Y_2|V_1}). \label{eq: UBe2}
\end{align}
where $\pi_{Y_2|V_2}$ and $\pi_{Y_2|V_1 V_2}$ are derived from (\ref{eq: InfoDivergUB}). The (\ref{eq: UBe2}) implies that as $n \rightarrow \infty$, $e_1 \rightarrow 0$. 

On the other hand, to upper bound $e_2$ in (\ref{eq: typicalset}) we need to use the Lemma \ref{lem: typicalPenalty}, since, $(\vv_1,\vv_2,\mathbf{y}_2) \in \mathcal{T}^n_\epsilon(P_{V_1V_2Y_2})$ defined on $P_{V_1,V_2, Y_2}$.
Therefore we have, 
\begin{align*} \nonumber
e_2 \leq&  \sum_{\substack{ \vv_1 \\ (\vv_1,\vv_2) \in \mathcal{T}^n_{\epsilon} }} P^n_{V_1}(\vv_1) \\
&\hspace{2cm} \log \bigg( \frac{2^{-n(H(Y_2|V_1,V_2) -\delta_1(\epsilon) ) } }{J_1 2^{-n(H(Y_2|V_2) + I(V_1;V_2)+ \delta_2(\epsilon))} } + 1 \bigg) \\
\leq&  \sum_{\substack{ \vv_1 \\ (\vv_1,\vv_2) \in \mathcal{T}^n_{\epsilon} }}2^{-n(H(V_1) + I(V_1;V_2)- \delta_3(\epsilon))} \\
& \hspace{2cm} \log \bigg( \frac{2^{-n(H(Y_2|V_1,V_2) -\delta_1(\epsilon) ) } }{J_1 2^{-n(H(Y_2|V_2) + I(V_1;V_2)+ \delta_2(\epsilon))} } + 1 \bigg) \\
\leq&  2^{n(H(V_1)+ \delta_4(\epsilon))}2^{-n(H(V_1) + I(V_1;V_2)- \delta_3(\epsilon))} \\
&\hspace{2cm}  \log \bigg( \frac{2^{-n(H(Y_2|V_1,V_2) -\delta_1(\epsilon) ) } }{J_1 2^{-n(H(Y_2|V_2) + I(V_1;V_2)+ \delta_2(\epsilon))} } + 1 \bigg) \\
\leq&  2^{-n( I(V_1;V_2)- \delta_3(\epsilon) -\delta_4(\epsilon))} \\
& \hspace{2cm}  \log \bigg( \frac{2^{-n(H(Y_2|V_1,V_2) -\delta_1(\epsilon) ) } }{J_1 2^{-n(H(Y_2|V_2) + I(V_1;V_2)+ \delta_2(\epsilon))} } + 1 \bigg)\\
\leq&  2^{-n( I(V_1;V_2)- \delta_3(\epsilon)-\delta_4(\epsilon))} \\
& \hspace{.001cm} \log \bigg( 2^{ -n(R'_1 + H(Y_2|V_1,V_2) - H(Y_2|V_2) - I(V_1;V_2) - \delta_1(\epsilon)-\delta_2(\epsilon) ) }+1 \bigg)\\
\leq & \log(e) 2^{-n(R'_1-I(Y_2;V_1|V_2) - \delta_5(\epsilon))},
\end{align*}
where, $\delta_5(\epsilon):= \delta_1(\epsilon)+\delta_2(\epsilon) + \delta_3(\epsilon) + \delta_4(\epsilon)$.
We have to impose $R'_1 > I(Y_2;V_1|V_2) + \delta_5(\epsilon)$ in order to assure $e_2 \rightarrow 0$ with $n \rightarrow \infty$. This condition, thus, evokes strong secrecy. 

Combining $R'_1 > I(Y_2;V_1|V_2)$, and symmetrically $R'_2 > I(Y_1;V_2|V_1)$, (\ref{eq: ratReleible1}) and (\ref{eq: ratReleible2}), we obtain the rate region as in the theorem 1. 

\end{IEEEproof} 



\section{Acknowledgement}
 This work is supported by the German Research Foundation (DFG) under Grant STA 864/7-1.


%
%
%
\bibliographystyle{IEEEtran}
\bibliography{myreferences}

\end{document}